\documentclass[aps,twocolumn,prmaterials,10pt, floatfix,superscriptaddress,final]{revtex4-2}
\pdfoutput=1
\usepackage[caption=false]{subfig}
\usepackage{graphicx, amsmath, amsfonts, amssymb, color}
\usepackage[%
  pdftex,
  breaklinks=true,
  colorlinks=true,
  linkcolor=blue,
  urlcolor=blue,
  citecolor=blue
]{hyperref}
\usepackage{verbatim}
\usepackage{xcolor}

\newcommand{\angstrom}{\mbox{\normalfont\AA}} 

\begin{document}

\title{Size-dependent bistability in multiferroic nanoparticles}
\author{Marc Allen}
\affiliation{Department of Physics and Astronomy, University of Victoria, Victoria, British Columbia V8W 2Y2, Canada}
\affiliation{Centre for Advanced Materials and Related Technology, University of Victoria, Victoria, British Columbia V8W 2Y2, Canada}
\author{Ian Aupiais}
\affiliation{Laboratoire Mat\'{e}riaux et Ph\'{e}nom\`{e}nes Quantiques, UMR 7162 CNRS, Universit\'{e} Paris Diderot, B\^{a}timent Condorcet 75205 Paris Cedex 13, France}
\author{Maximilien Cazayous}
\affiliation{Laboratoire Mat\'{e}riaux et Ph\'{e}nom\`{e}nes Quantiques, UMR 7162 CNRS, Universit\'{e} Paris Diderot, B\^{a}timent Condorcet 75205 Paris Cedex 13, France}
\author{Rog\'{e}rio \surname{de Sousa}}
\affiliation{Department of Physics and Astronomy, University of Victoria, Victoria, British Columbia V8W 2Y2, Canada}
\affiliation{Centre for Advanced Materials and Related Technology, University of Victoria, Victoria, British Columbia V8W 2Y2, Canada}

\begin{abstract}
Most multiferroic materials with coexisting ferroelectric and magnetic order exhibit cycloidal antiferromagnetism with wavelength of several nanometers. The prototypical example is bismuth ferrite (BiFeO$_3$ or BFO), a room-temperature multiferroic considered for a number of technological applications. While most applications require small sizes such as nanoparticles, little is known about the state of these materials when their sizes are comparable to the cycloid wavelength. This work describes a microscopic theory of cycloidal magnetism in nanoparticles based on Hamiltonian calculations. It is demonstrated that magnetic anisotropy close to the surface has a huge impact on the multiferroic ground state. For certain nanoparticle sizes the modulus of the ferromagnetic and ferroelectric moments are bistable, an effect that may be used in the design of ideal memory bits that can be switched electrically and read out magnetically. 
\end{abstract}



\maketitle

\section{Introduction}

Multiferroic materials display coexisting ferroic orders at the same temperature \cite{spaldin_multiferroics:_2010}. An important class is the magnetoelectric multiferroics, with coexisting ferroelectricity and magnetism, usually antiferromagnetism \cite{de_sousa_holy_2016}. The impact of ferroelectricity on the magnetic state occurs through the spin-orbit interaction, more specifically through the spin-current contribution of the Dzyaloshinskii-Moryia (DM) interaction. This interaction induces spiral magnetism of the cycloidal type, with cycloid period $\lambda=2\pi/Q$ much larger and incommensurate with the material's lattice spacing $a$ \cite{lebeugle_electric-field-induced_2008, matsuda_magnetic_2012, rahmedov_magnetic_2012}. Conversely, cycloidal magnetism induces ferroelectricity \cite{katsura_spin_2005, mostovoy_ferroelectricity_2006}.

A notable example of magnetoelectric multiferroic is bismuth ferrite ({BiFeO}$_3$ or BFO), one of the few room-temperature multiferroics~\cite{catalan_physics_2009, park_structure_2014} with potential for technological applications such as electrically-written magnetic memories \cite{scott_applications_2007, heron_electric-field-induced_2011, wu_full_2013, heron_deterministic_2014}
or photoelectricity \cite{yang_photovoltaic_2009}. Bulk BFO is ferroelectric at temperatures below 1100~K, with a record high polarization $P\approx 100~\mu\mathrm{C/cm}^2$ at room temperature. Below 640~K the Fe spins form a nearly cubic antiferromagnetic lattice with cycloidal spin ordering of period equal to $\lambda_{\rm Bulk}=630$~\angstrom~\cite{sosnowska_spiral_1982, ramazanoglu_temperature-dependent_2011}, much larger than the lattice parameter $a=3.96$~\angstrom. 

Memory applications require miniaturized multifunctional devices with multiferroic size approaching $\lambda_{\rm Bulk}$. So far only a few studies have considered the impact of finite size on BFO's  ferroelectric and magnetic properties. It was shown experimentally that BFO nanoparticles remain ferroelectric and antiferromagnetic at room temperature, but with decreased Curie and N\'{e}el transition temperatures, and 
enhanced ferromagnetism \cite{mazumder_particle_2006, park_size-dependent_2007, selbach_size-dependent_2007, chen_size-dependent_2010, AnnapuReddy2012, huang_peculiar_2013,carranza-celis_control_2019}. 

The presence of a depolarizing electric field in finite-sized ferroelectrics is known to reduce their Curie temperature and polarization $P$. There exists a critical size below which the nanoparticle ceases to be ferroelectric \cite{Shih1994, ren_size_2010}. This effect was measured in free-standing BFO nanoparticles \cite{selbach_size-dependent_2007}, where it was shown that 
$P\approx P_{\rm Bulk}$ for sizes down to $30$~nm, with $P$ reduced to 
$0.75P_{{\rm Bulk}}$ for size $13$~nm. Extrapolating to even smaller nanoparticles suggested a critical size of approximately $9$~nm.

Demagnetizing fields play a similar role in ferromagnetic nanoparticles, favouring the formation of magnetic vortex states \cite{Sheng2017}. However, vortex states do not occur in antiferromagnetic nanoparticles,
even when they have a weak ferromagnetic moment arising from spin canting or uncompensated spins at the surface. 
The small ferromagnetic moment leads to a demagnetizing energy that is several orders of magnitude smaller than the antiferromagnetic exchange energy. As a result, the spin-spin dipolar interaction can be neglected in models for antiferromagnetic nanoparticles \cite{kodama_finite_1997}. However, the impact of finite size and the role of surface interactions on the spin texture of cycloidal multiferroics has not yet been explored.

Nanoparticles differ qualitatively from the bulk due to their larger surface-to-volume ratios. Here it is argued that the magnetic order of multiferroic nanoparticles is greatly influenced by magnetic interactions at the surface. The most important of these interactions is single-ion anisotropy \cite{sosnowska_origin_1995, fishman_spin_2013}, which originates
from two large spin-orbit contributions of opposite sign, both associated to the location of the Bi ion in BFO \cite{de_sousa_theory_2013, weingart_noncollinear_2012}. 
At the surface of the nanoparticle reduced symmetry means that the two contributions to single-ion anisotropy no longer cancel each other out, leading to large magnetic anisotropy at or nearby the surface. 
Reduced symmetry at the surface has also been thought to increase surface anisotropy due to factors including broken exchange bonds and interatomic distance variation \cite{kodama_magnetic_1999, winkler_surface_2005}. 

Strain also increases single-ion anisotropy and this explains why the magnetic cycloid order is destroyed in BFO thin films grown on top of substrates with large relative strain \cite{Sando2013}. The present research article focuses on unstrained nanoparticles, i.e., those that are either freestanding or grown on top of a lattice-matched substrate. Experiments show that substrates with relative strain smaller than $0.5$\% preserve the cycloid order of BFO \cite{Sando2013}. Below it is shown that in these unstrained nanoparticles the combination of cycloidal spin order and surface magnetic anisotropy leads to \emph{multiferroic bistability}.

\section{Model for multiferroic nanoparticles}

In this article a model for the impact of surface anisotropy on a multiferroic nanoparticle is proposed. As a starting point, consider the Hamiltonian describing antiferromagnetism in a magnetoelectric multiferroic \cite{katsura_spin_2005, matsuda_magnetic_2012, rahmedov_magnetic_2012},
\begin{equation}
{\cal H}_0 = \frac{1}{2}\sum_{i,\mathbf{\hat{v}}} \left[J \mathbf{S}_i\cdot\mathbf{S}_{i+\mathbf{\hat{v}}} +D \mathbf{\hat{P}}\cdot \mathbf{\hat{v}}\times \left( \mathbf{S}_i\times\mathbf{S}_{i+\mathbf{\hat{v}}}\right)\right].
\label{eq:H0}
\end{equation}
The classical vectors $\mathbf{S}_i = (S_{ix},S_{iy},S_{iz})$ represent the $i^{\mathrm{th}}$ spin in a hypercubic lattice (dimension $d=1,2,3$) \emph{without} periodic boundary conditions. 
For example, the $d=2$ case has each spin located at $\mathbf{R}_i=a(i_x,i_y)$, with each $i_{\alpha}=1,\ldots, N$, etc ($\alpha=x,y, z$ and total number of spins is $N^d$).
The unit vectors $\mathbf{\hat{v}}$ link nearest neighbours coupled by exchange energy $J>0$ (antiferromagnetism).

The spins are affected by the ferroelectric moment $\mathbf{P}$ via the spin-current energy $D$. The second term in Eq.~(\ref{eq:H0}) can be interpreted as $-a^3 \mathbf{P}\cdot \mathbf{E}^{\rm spin}_{i}$, with 
$\mathbf{E}^{\rm spin}_{i}$ the spin-induced electric field at site $\mathbf{R}_i$. This local field changes the polarization according to $\Delta \mathbf{P}_i=\chi \mathbf{E}^{\rm spin}_i$, with $\chi$ an electric susceptibility \cite{lee_giant_2015}, which averaged over all sites leads to the \emph{spin-induced ferroelectric moment} 
\begin{equation}
\mathbf{P}_{\rm spin}=-\frac{D\chi}{2Pa^3N^d} \sum_{i,\mathbf{\hat{v}}} \mathbf{\hat{v}}\times \left( \mathbf{S}_i\times\mathbf{S}_{i+\mathbf{\hat{v}}}\right).
\label{eq:P_spin}
\end{equation}

The ground state of Hamiltonian~(\ref{eq:H0}) occurs for spins lying in the $\mathbf{\hat{Q}}\mathbf{\hat{P}}$ plane,
\begin{equation}
	\label{eq:harm_spin}
	\mathbf{S}_{i} = \left ( -1 \right )^{\sum_\alpha i_\alpha} 
	\left ( \sin(\phi_{i}) \, \mathbf{\hat{Q}} + \cos(\phi_{i}) \, \mathbf{\hat{P}} \right), 
\end{equation}
with cycloid unit vector $\mathbf{\hat{Q}}$ simultaneously perpendicular to $\mathbf{\hat{P}}$ and parallel to one of the nearest neighbour directions $\mathbf{\hat{v}}$. When $N\rightarrow \infty$ (the bulk limit) the angle $\phi_i$ is simply given by  \cite{lebeugle_electric-field-induced_2008, matsuda_magnetic_2012, rahmedov_magnetic_2012}
\begin{equation}
	\phi_{i} = \phi_{0} + \mathbf{Q}\cdot \mathbf{R}_i,
	\label{eq:phi}
\end{equation}
with $\phi_0$ an arbitrary phase slip, and $\mathbf{Q}$ a constant 
cycloid wavevector with $|\mathbf{Q}|=Q_{{\rm Bulk}}=\arctan{(D/J)}/a$. Such a state has $\mathbf{P}_{\rm spin}\propto -\sin{(Q_{\rm Bulk}a)}\mathbf{\hat{P}}$. 

Consider the Hamiltonian for surface anisotropy,
\begin{equation}
{\cal H}_S=-K_S \sum_{i\in {\rm surfaces}} \left(\mathbf{S}_i\cdot \mathbf{\hat{n}}\right)^{2}, 
\label{eq:HS}
\end{equation}
where $K_S$ is the extra anisotropy energy arising due to the reduced symmetry either at the nanoparticle/air surface or the nanoparticle/substrate interface (in case the nanoparticle is on top of a substrate). The surface unit vector $\mathbf{\hat{n}}$ points perpendicular to the surface, with spins lying at the intersection of $n'$ surfaces appearing $n'$ times in the sum. There is an important difference between the surfaces with $\mathbf{\hat{n}}\parallel\mathbf{\hat{P}}$ and the ones that are not. The former necessarily has $\mathbf{Q}\perp \mathbf{\hat{n}}$, so that the surface is made up of cycloid chains, with \emph{all} spins subject to anisotropy. In this case, $|K_S|>0$ greatly reduces the surface $Q$ and the cycloid is destroyed ($Q=0$) for $|K_S|>D^2/J$ \cite{buhot_driving_2015}. This reduction in $Q$ propagates a distance close to $\lambda_{\rm Bulk}$ towards the interior of the nanoparticle due to the \emph{proximity effect}. 
For surfaces with $\mathbf{\hat{n}}\neq\pm\mathbf{\hat{P}}$, $K_S$ affects only the edge of the cycloid chains penetrating into the material. These chains can adjust their $Q$ to minimize the impact of surface anisotropy as it is shown below. This will be referred to as the \emph{edge effect}.

\section{The edge effect}

\begin{figure}	
	\centering
	\includegraphics[width=0.49\textwidth]{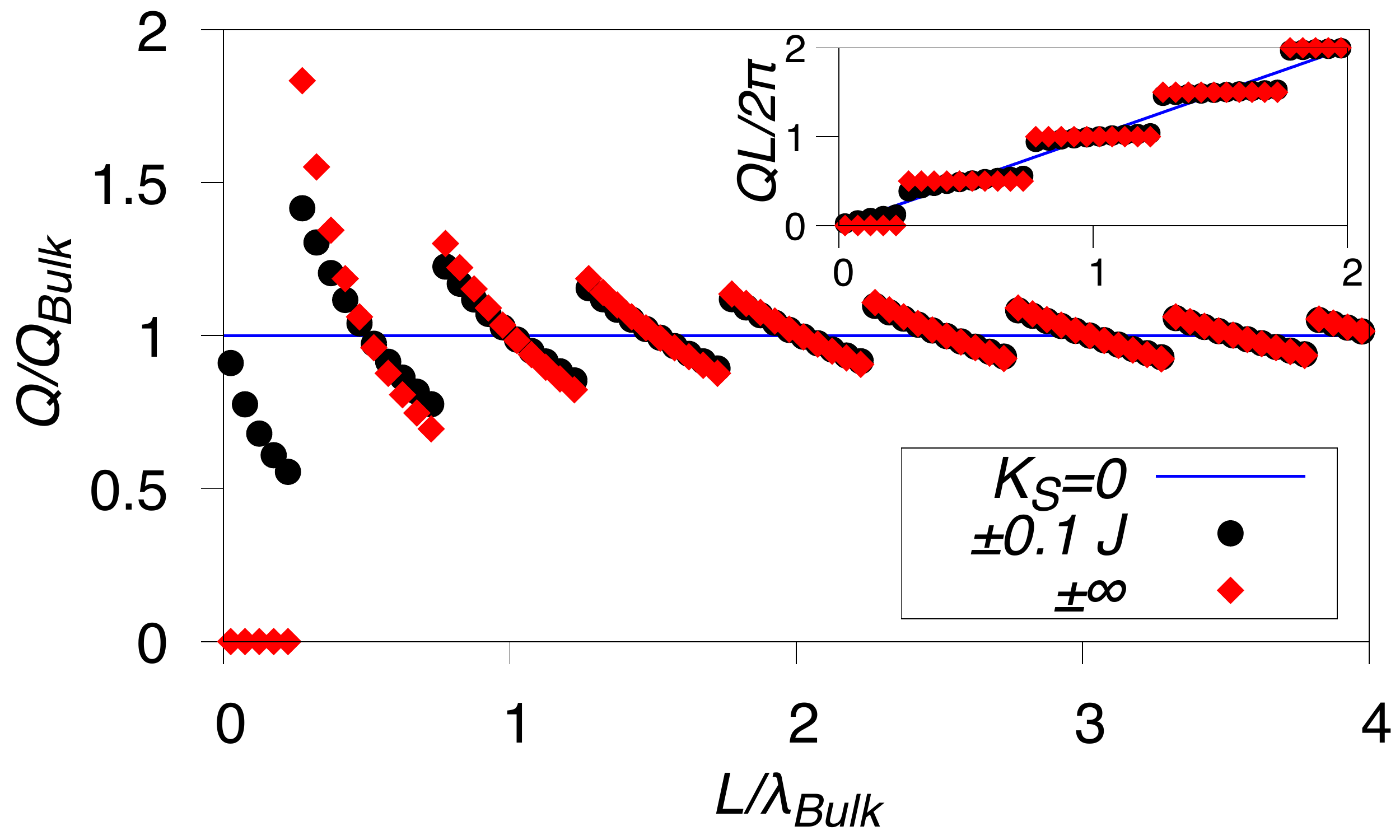}
	\caption{(colour online) Ground state cycloid wavevector $Q$ as a function of system size $L$ for the one-dimensional model with edge anisotropy $K_S/J=0, \pm 0.1, \pm\infty$. The quantities $Q$ and $L$ are normalized by $Q_{\rm Bulk}$ and $\lambda_{\rm Bulk}$, where $Q_{\rm Bulk}=2\pi/\lambda_{\rm Bulk}$ is the cycloid wave vector for the infinite system. Inset: Winding number $QL/2\pi$ versus $L/\lambda_{{\rm Bulk}}$.}
	\label{fig:1dks}
\end{figure}

Consider a spin chain along $x$ with total length equal $L=(N-1)a$, and take $\mathbf{\hat{P}}=\mathbf{\hat{z}}$. 
The edges of the chain described by $i=1$ and $i=N$ are the only ones subject to surface anisotropy in Eq.~(\ref{eq:HS}). The total Hamiltonian ${\cal H}_0+{\cal H}_S$ was minimized numerically for each size $N$, with $\lambda_{{\rm Bulk}}/a=40$ remaining fixed (corresponds to $D/J=0.15708$). The numerical minimization was done using the Nelder-Mead method with several random starting points (NMinimize function in \emph{Mathematica}). The energy minima has the form of Eq.~(\ref{eq:phi}) with $\phi_0$ pinned to certain fixed values and 
$\mathbf{Q}=Q\mathbf{\hat{x}}$ with $Q$ strongly dependent on size $L$. Figure~\ref{fig:1dks} shows the results of the minimization for $Q/Q_{{\rm Bulk}}$ versus $L/\lambda_{{\rm Bulk}}$, for $N$ even and surface anisotropy $K_S/J=0,~\pm0.1,~\pm\infty$. The results for easy axis ($K_S>0$) were identical to the ones for easy plane ($K_S<0$). When $K_S=0$ the $Q$ was independent of size $L$ and equal to $Q_{{\rm Bulk}}$, as expected. The introduction of edge anisotropy caused asymptotic behaviour in $Q$. Its value became proportional to $1/L$ with jump discontinuities at $L_n = (2n + 1)\lambda_{\mathrm{Bulk}}/4$ ($n = 0, 1, 2, \ldots$). 

The origin of this behavior is the necessity to make the two edge spins perpendicular ($K_S>0$) or parallel ($K_S<0$) to the surface, at the same time that the angle between every spin is kept constant. This can only be achieved by increasing or decreasing the angle between each spin to accommodate an integer multiple of $\pi/Q=\lambda/2$.  More insight is gained by looking at the winding number $QL/(2\pi)$: This is the number of $2\pi$ revolutions inside the chain. The inset of Fig.~\ref{fig:1dks} shows that winding number versus $L$ has well-defined plateaus at half-integer values.

At the locations of the discontinuities ($L=L_n$), the equilibrium value of $Q$ is bistable -- the energy landscape is a double well with global minima at two different values of $Q$. 
Later it will be shown that this bistability in $Q$ implies bistability of the modulus of the total electric and magnetic moments of the nanoparticle. 

The drastic variability of $Q$ as a function of size shown in Fig.~\ref{fig:1dks} can be directly observed in experiments probing the cycloid in an ensemble of nanoparticles. A direct probe of $Q$ is to measure the cycloidal magnons using Raman \cite{cazayous_possible_2008} or TeraHertz spectroscopy \cite{nagel_terahertz_2013}. The cycloidal magnons lead to optical resonances at frequencies approximately proportional to integer multiples of Q \cite{de_sousa_optical_2008}. For an ensemble of nanoparticles, the variability in $Q$ will lead to inhomogeneous broadening of these optical resonances. 

\section{Competition between the edge and proximity effects} 

To see what happens in the presence of four surfaces (two sides with $\mathbf{\hat{n}}=\pm\mathbf{\hat{x}}\perp\mathbf{\hat{P}}$ and top/bottom with $\mathbf{\hat{n}}=\pm\mathbf{\hat{z}}\parallel\mathbf{\hat{P}}$) consider a $d=2$ platelet of spins oriented along the $xz$ direction, with spin labels $i=(i_x, i_z)$. 
As a check, the first calculation was done with $K_S\neq 0$ only for spins located on the side surfaces, at $i=(1,i_z)$ and $i=(N,i_z)$. 
The optimal $Q$ values were all equal for different $i_z$'s. With no surface anisotropy along the top and bottom of the platelet, the spins behaved as in the $d=1$ case, as expected.

\begin{figure}
	\centering
	\includegraphics[width=0.49\textwidth]{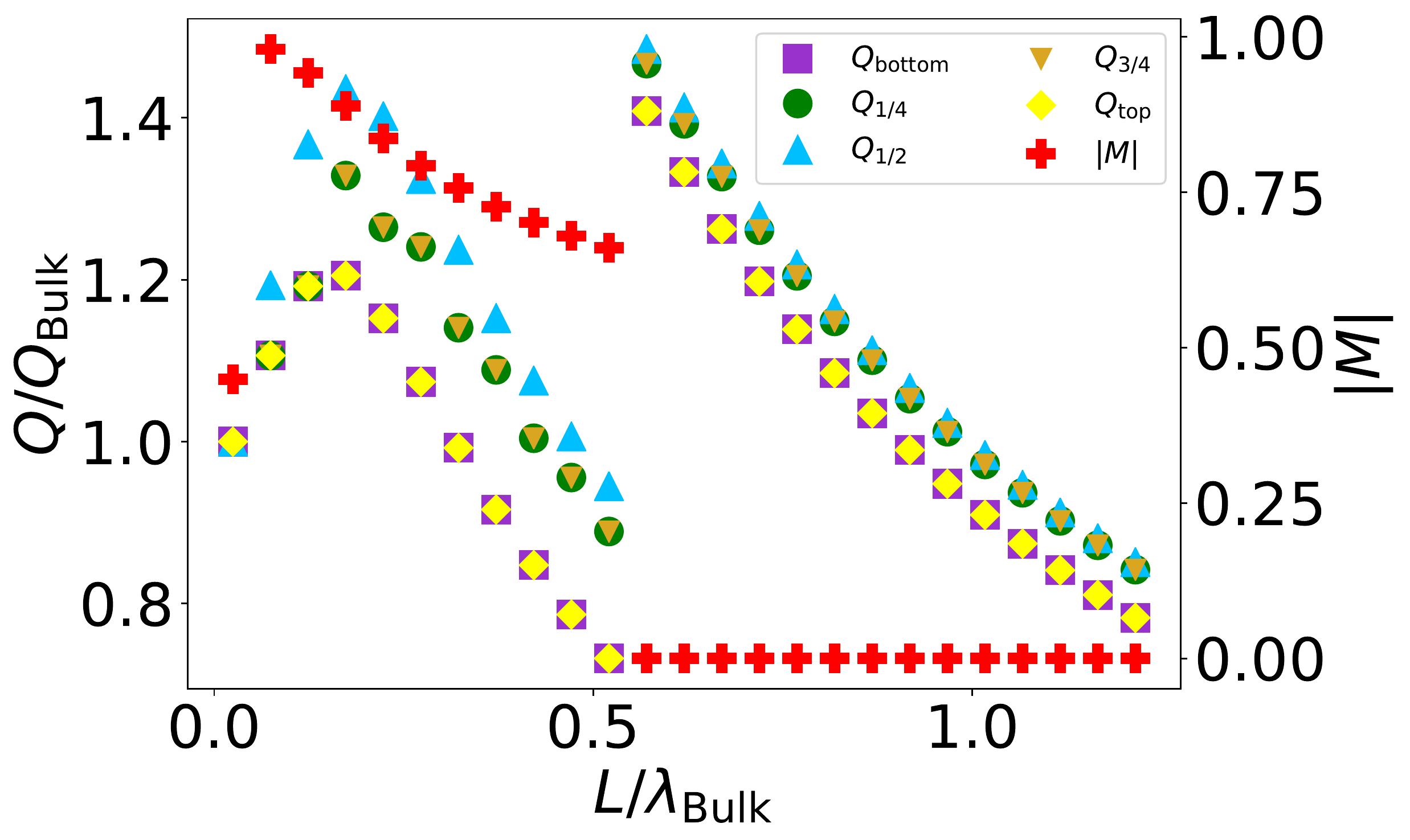}
	\caption{(colour online) Cycloid wavevector $Q$ versus length for a compensated two-dimensional platelet of side $L$, oriented along the $\mathbf{\hat{Q}}\mathbf{\hat{P}}$ plane (xz), with $K_S/J = -0.1$. 
The values of $Q$ depend on the spin location along $z$; the subscripts in $Q$ indicate the value of z, e.g. $Q_{3/4}$ corresponds to $z=3L/4$, etc. At $L/\lambda_{\rm Bulk}\approx 0.5$ there is a jump discontinuity and the values of $Q$ are bistable. Also shown is the value of ferromagnetic moment $|\mathbf{M}|$ calculated from Eq.~(\ref{eq:M}), in units of $D'/J$. Note how $|\mathbf{M}|$ scales proportionally to $Q$ for $L< 0.5 \lambda_{\rm Bulk}$.}
	\label{fig:q_ks-0p1}
\end{figure}

Including surface anisotropy along all four surfaces made the cycloid anharmonic, in that the optimal $Q$ depended on the index $i_z$. 
Figure~\ref{fig:q_ks-0p1} shows results for $N$ even (compensated) and $K_S/J=0.1$ (easy axis) on all surfaces. The values of $Q$ are listed according to their position with respect to the $z$-direction, e.g. $Q_{{\rm bottom}}$ is for the bottom row ($i_z=1$) and $Q_{1/2}$ is for the middle row ($i_z=N/2$). Note the mirror symmetry (e.g. $Q_{{\rm top}}=Q_{{\rm bottom}}$, and $Q_{1/4}=Q_{3/4}$), and 
how the value of $Q$ increases gradually as the spin location moves towards the centre. The competing interactions impose a ``proximity effect'' for surface anisotropy that affects spins well into the centre of the platelet.

Remarkably, for $L\approx 0.5\lambda_{\mathrm{Bulk}}$ all $Q$'s become bistable in the $K_S/J=-0.1$ case (note the jump discontinuity in Fig.~\ref{fig:q_ks-0p1}). This is a surprising result, in view of the fact that the presence of several different wavelengths $\lambda(z)=2\pi/Q(z)$ does not allow the fitting of odd integer multiples of  a single $\lambda/4$ inside $L$. The bistability occurs for several other values of surface anisotropy. For example, when $K_S/J=0.1$ the bistability occurs at $L\approx 0.8\lambda_{{\rm Bulk}}$ and $1.2\lambda_{\mathrm{Bulk}}$. The results for $N$ odd (uncompensated) were quite similar, with the bistability happening at a slightly different $L$. The minimum energy configuration had  unpaired spins in each chain aligning antiparallel to each other, 
leading to additional contribution to the ferromagnetic moment per spin $|\mathbf{M}|$ approximately equal to $1/L^2$, quite similar to the unpaired moment in non-cycloidal antiferromagnets  in two and three dimensions \cite{kodama_finite_1997}.

\begin{figure*}[t]
\begin{center}
\subfloat[Spin-induced ferroelectric moment.]{\includegraphics[width=0.49\textwidth]{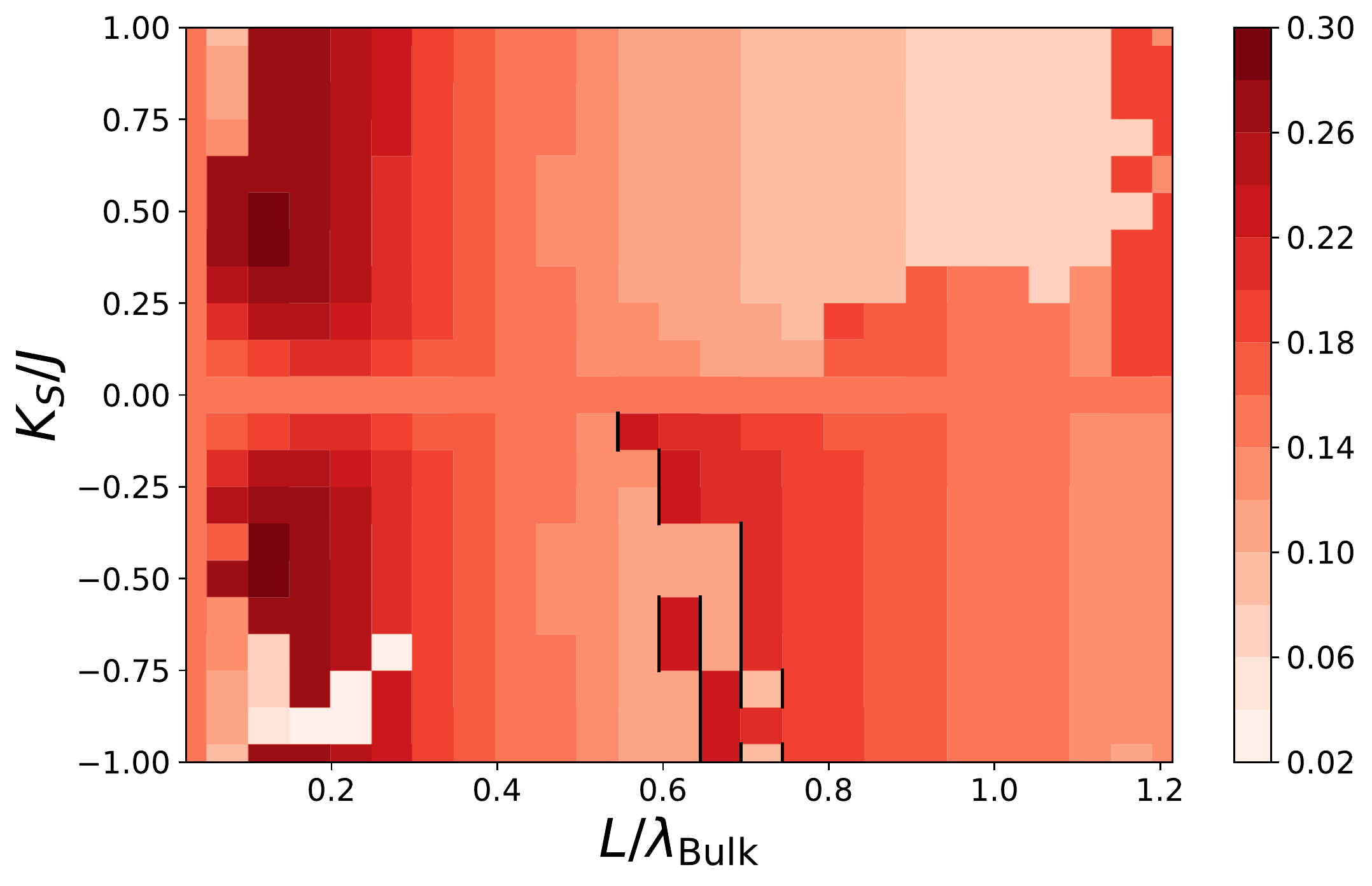}
\label{fig:electricfield_compensated}}
\subfloat[Spin-canting-induced ferromagnetic moment.]{\includegraphics[width=0.49\textwidth]{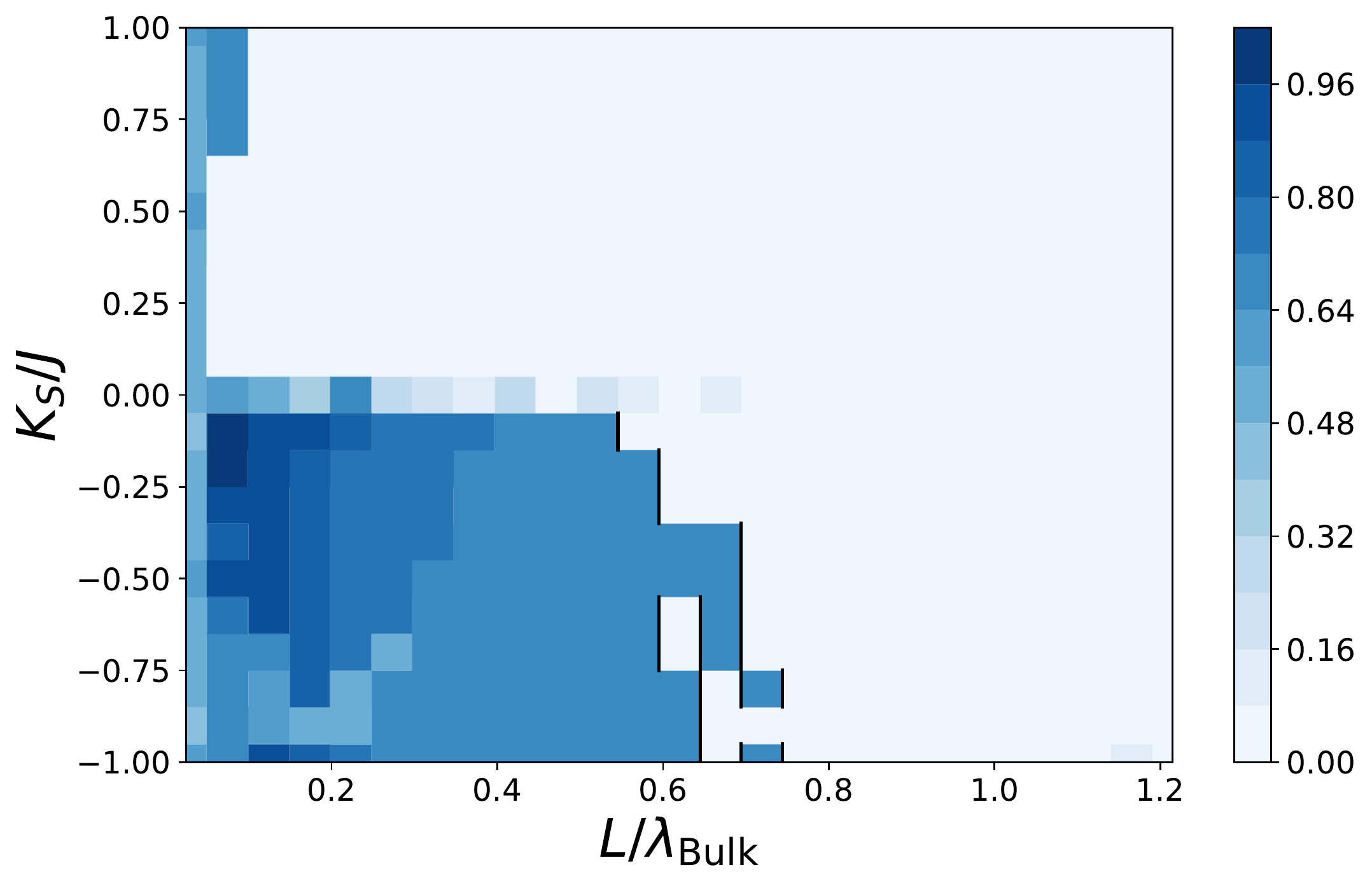}
\label{fig:ferromagnetism_compensated}}
\end{center}
\caption{(colour online) Phase diagram for the ferroelectric and ferromagnetic properties of a compensated $L\times L$ nanoparticle platelet, as a function of size $L$ and surface anisotropy $K_S$. (a) Ferroelectric polarization per spin $|\mathbf{P}_{\rm spin}|$ calculated from Eq.~(\ref{eq:P_spin}), in units of $D\chi/(Pa^3)$. (b) Ferromagnetic moment per spin $|\mathbf{M}|$ calculated from Eq.~(\ref{eq:M}), in units of $D'/J$. The black lines mark jump discontinuities in both $|\mathbf{P}_{\rm spin}|$ and $|\mathbf{M}|$. 
At these points the nanoparticle is bistable with respect to both electric and magnetic properties. 
\label{fig:electric_magnetic}}
\end{figure*}

The dependence of $Q$ on nanoparticle size and $K_S$ is depicted in Fig.~\ref{fig:electricfield_compensated}, the phase diagram for the modulus of the spin-induced ferroelectric moment $|\mathbf{P}_{\rm spin}|$ which is proportional to $\langle \sin{(Qa)}\rangle$.

\section{Ferromagnetism in multiferroic nanoparticles} 

The reduction of $Q$ towards the surfaces $\mathbf{\hat{n}}\parallel\mathbf{\hat{P}}$ dramatically impact the ferromagnetism of nanoparticles. To see this, 
the spin-canting contribution of the DM interaction is considered, 
\begin{equation}
{\cal H}_{{\rm DM}} =\frac{D'}{2}\sum_{i,\mathbf{v}}(-1)^{\sum_\alpha i_\alpha} \mathbf{\hat{z}}\cdot \mathbf{S}_i\times \mathbf{S}_{i+\mathbf{v}}.
\label{eq:HDM}
\end{equation}
For $D'\ll J$ this interaction results in nanoparticle ferromagnetic moment per spin given by 
\begin{equation}
    \mathbf{M}=\frac{1}{N^dS}\sum_{i}\mathbf{S}_i=\frac{D'}{2JN^dS} \sum_{i}(-1)^{\sum_\alpha i_\alpha} \mathbf{\hat{z}}\times\mathbf{S}_{i}.  
\label{eq:M}
\end{equation}

For bulk cycloids the argument of the sum in Eq.~(\ref{eq:M}) is a sine wave with period $\lambda_{\rm Bulk}$ pointing perpendicular to the cycloid plane. Thus $\mathbf{M}$ averages out over distances $L\gg \lambda_{\rm Bulk}$ \cite{ramazanoglu_local_2011}. In nanoparticles with size $L\lesssim\lambda_{\rm Bulk}$ the ferromagnetic moment does not average out. Measured in several experiments, it was interpreted to arise from uncompensated  antiferromagnetism \cite{mazumder_particle_2006, park_size-dependent_2007, selbach_size-dependent_2007, chen_size-dependent_2010, AnnapuReddy2012, huang_peculiar_2013, carranza-celis_control_2019}.

Figure~\ref{fig:ferromagnetism_compensated} shows the phase diagram for $|\mathbf{M}|$ in units of $D'/J$ for compensated samples. Quite remarkably, the nanoparticles have sizable ferromagnetism in a large parameter range. Here the ferromagnetism arises close to the surfaces perpendicular to $\mathbf{\hat{P}}$, where $Q$ is greatly reduced so the spin canting contribution to $\mathbf{M}$ does not average out. This result shows that 
spin canting at the surface provides an additional mechanism for nanoparticle surface ferromagnetism, scaling as $|\mathbf{M}|\sim 1/L$ in three dimensions, in agreement with experiments \cite{park_size-dependent_2007}. Even for small values of $D'/J$ this can be much larger than the moment arising from unpaired spins in uncompensated surfaces \cite{kodama_finite_1997}.

Figure~\ref{fig:electric_magnetic} demonstrates a rich magnetoelectric phase diagram as a function of particle size and surface anisotropy. There are several lines of bistability for $\mathbf{P}_{\rm spin}$. 
Each time this happens, there are four possible states for $\mathbf{M}$ ($\pm \mathbf{M}$ with $|\mathbf{M}|$ assuming two different values). 

The bistability in $\mathbf{P}_{\rm spin}$ and $\mathbf{M}$ can be used as a memory where either $\mathbf{P}_{{\rm spin}}$ or $\mathbf{M}$ encodes information. With $\mathbf{P}_{\rm spin}\sim 3$~$\mu$C/cm$^2$ \cite{lee_giant_2015} the bit is switchable electrically with electric fields of the order of $10^2$~V/cm (see endnote 27 in \cite{de_sousa_theory_2013}). With $M\sim 0.1$~$\mu_B/$Fe \cite{ramazanoglu_local_2011} corresponding to a local field of 200~G, it can be read out magnetically using usual hard drive read heads, or with state of the art optical read heads based on diamond NV-center magnetometry \cite{Gross2017}. Altogether such a memory bit corresponds to the ``ideal memory'' that is electrically written and magnetically read envisioned in \cite{scott_applications_2007}. 

\section{Conclusions} 

The considerations above allows general predictions about the magnetoelectric behaviour of nanoparticles of arbitrary shape. For example, in $d=3$ consider a nanoparticle shaped as a cube with surfaces perpendicular to the $x, y, z$ axes. For $K_S<0$ (easy plane) the spin configuration that minimizes energy
consists of $xz$ spin planes stacked next to each other, each with spin configuration identical to the $d=2$ platelets described in Fig.~\ref{fig:electric_magnetic}. Note that such configuration is a planar cycloid with $\mathbf{Q}=Q(z)\mathbf{\hat{x}}$, leading to anisotropy energy equal to zero for the two surfaces with $\mathbf{\hat{n}}=\pm \mathbf{\hat{y}}$. 

The case of $K_S>0$ (easy axis) in $d=3$ warrants additional calculations. Minimizing the $\mathbf{\hat{n}}=\pm\mathbf{\hat{x}}, \pm \mathbf{\hat{y}}$
surface energies leads to a twist configuration for the $\mathbf{Q}$ vector. Consider a cylindrical geometry with radius $R$ in the xy plane and axis length $L_z\rightarrow \infty$. Numerical minimization with $\mathbf{Q}=Q\mathbf{\hat{\rho}}$ (the ``Q-monopole'') shows that $Q(\rho)\ll Q_{\rm Bulk}$ for $\rho\lesssim \lambda_{\rm Bulk}$ and $Q(\rho)\approx Q_{\rm Bulk}$ for $\rho\gtrsim \lambda_{\rm Bulk}$, with jump discontinuities in $Q$ at $R=(2n+1)\lambda_{\rm Bulk}/4$ due to the edge effect. For $L_z$ finite the $Q's$ are further reduced due to the proximity effect of the $\mathbf{\hat{n}}=\pm\mathbf{\hat{z}}$ surfaces. 

These considerations show that nanoparticles of arbitrary shape with $K_S>0$ will have smaller $\mathbf{P}_{\rm spin}$ and larger $\mathbf{M}$ than shown in Fig.~\ref{fig:electric_magnetic} within an interior volume $\lesssim \lambda_{\rm Bulk}^{3}$. Outside this volume the $\mathbf{Q}$ will twist so that it stays perpendicular to the surface. 

In summary, the spin texture of multiferroic nanoparticles was studied with numerical calculations. Surface anisotropy was shown to greatly impact the value of the cycloid wavevector $Q$, the spin-induced ferroelectric moment $\mathbf{P}_{\rm spin}$, and ferromagnetic moment $\mathbf{M}$. 
A rich magnetoelectric phase diagram comes out as a function of size and surface anisotropy with ferroelectric and ferromagnetic bistable points. The size-dependent bistability phenomena represents exciting prospects for the design of multifunctional memories using multiferroic nanoparticles.

\begin{acknowledgments}
This work was supported by NSERC (Canada) through its Discovery program (RGPIN-2015- 03938). The authors thank T. R\~{o}\~{o}m for critical reading of the manuscript. 
\end{acknowledgments}

\bibliography{surface_anisotropy}

\end{document}